\documentclass[twocolumn,showpacs,preprintnumbers,amsmath,amssymb]{revtex4}
\usepackage{graphicx}% Include figure files
\usepackage{dcolumn}% Align table columns on decimal point
\usepackage{bm}% bold math
\input{epsf.tex}

\begin{document}

\title{$\eta$ Photoproduction on the proton revisited: \\
Evidence for a narrow N(1688) resonance.}

\author{V.~Kuznetsov$^{1,2}$\footnote{E-mail SlavaK@jlab.org, Slava@cpc.inr.ac.ru}}
\author{M. V.~Polyakov$^{3,4}$\footnote{E-mail maxim.polyakov@tp2.ruhr-uni-bochum.de}}
\author{T.~Boiko$^5$}
\author{J.~Jang$^1$}
\author{A.~Kim$^1$}
\author{W.~Kim$^1$}
\author{A.~Ni$^1$}

\affiliation{$^1$Kyungpook National University, 702-701, Daegu, Republic of Korea}
\affiliation{$^2$Institute for Nuclear Research, 117312, Moscow, Russia}
\affiliation{$^3$Institute f\"ur Theoretische Physik II, Ruhr-Universit\"at Bochum,
D - 44780 Bochum, Germany,}
\affiliation{$^4$St. Petersburg Institute for Nuclear Physics, Gatchina, 188300, St. Peterburg, Russia,}
\affiliation{$^5$Belarussian State University, 220030, Minsk, Republic of Belarus}

\date{\today}

\begin{abstract}

Revised analysis of $\Sigma$ beam asymmetry for
$\eta$ photoproduction on the free proton reveals
a resonant structure at $W\sim 1.69$ GeV. Comparison of experimental
data with multipole decomposition based on the E429 solution
of the SAID partial wave analysis and including narrow
states, suggests a narrow ($\Gamma \leq 15$ MeV) resonance.
Possible candidates are $P_{11}$, $P_{13}$, or $D_{13}$ resonances.
The result is considered in conjunction with the recent
evidence for a bump-like structure at $W\sim 1.67 - 1.68$ GeV in
quasi-free $\eta$ photoproduction on the neutron.

\end{abstract}

\maketitle

\begin{figure}
\vspace*{-0.4cm}
\centerline{\epsfverbosetrue\epsfxsize=8.5cm\epsfysize=6.5cm\epsfbox{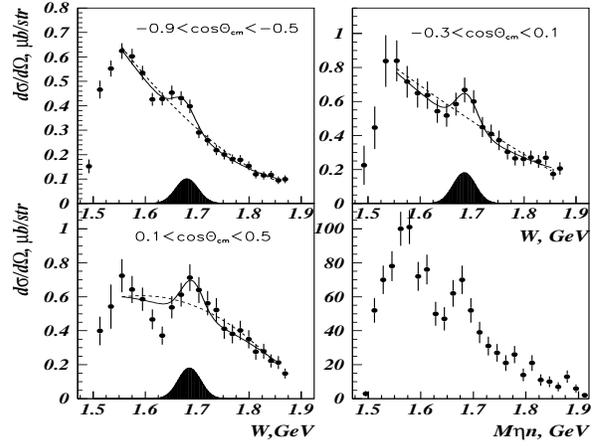}}
\caption{Quasi-free cross sections and $\eta n$ invariant mass spectrum
for the $\gamma n \to \eta n$ reaction (data from \protect\cite{gra1}).
Solid lines are the fit by the sum of 3-order polynomial and
narrow state. Dashed lines are the fit by 3-order polynomial only.
Dark areas show the simulated signal of a narrow state.}
\vspace*{-0.3cm}
\label{fig:etan}
\vspace{-0.3cm}
\end{figure}

Experimental study of the quasi-free $\gamma n \to \eta n$
reaction at GRAAL\cite{gra1}, CB/TAPS@ELSA\cite{kru}, and LNS-Tohoku\cite{kas}
facilities provided the evidence for a relatively narrow
resonant structure at  $W\sim 1.67 - 1.68$ GeV. The structure has been observed
as a bump in the quasi-free cross section and in the $\eta n$ invariant
mass spectrum (Fig.~\ref{fig:etan}).

The width of the bump in the quasi-free cross section was found to be close
to that expected due to smearing by Fermi motion of the target neutron
bound in the deuteron. A narrow resonance which would manifest
as a peak in the cross section on the free neutron,
would appear in the quasi-free cross section as a bump of
about 50~MeV width~\cite{gra1}. The simulated signal of such resonance
with the mass $M=1680$ MeV and the width $\Gamma=10$ MeV is shown
in Fig.~\ref{fig:etan}. The cross section is well
fitted by the sum of a third-order polynomial and the contribution
of the resonance.

The width of the peak in the $\eta n$ invariant mass spectrum is about $40$~MeV.
This quantity is much less affected by Fermi motion but includes large
uncertainties due to detector response. The width of the peak is close
to the instrumental resolution~\cite{gra1}.

Such bump is not seen in $\eta$ photoproduction on the proton. The cross section
on the proton exhibits only a minor peculiarity in this mass region~\cite{etap}.
Therefore the bump in $\eta$ photoproduction on the neutron
may be a manifestation of a nucleon resonance with unusual properties:
the possibly narrow width and the much stronger photocoupling to the neutron
than to the proton. Its identification is now a challenge
for both theory and experiment.

This task is particularly important because the properties of
the possible underlying resonance well coincide with the predictions
for the second $P_{11}$ member of the anti-decuplet
of exotic baryons(pentaquarks)~\cite{dia}.
This particle (if it exists) has the strangeness $S=0$.
Its mass is expected to be in the range $1.65-1.7$ GeV~\cite{dia1,str},
with a total width of about 10-15 MeV~\cite{str}.
A benchmark signature of the non-strange pentaquark could be
its photoproduction on the nucleon. The chiral soliton model predicts
that photoexcitation of the non-strange pentaquark has to be considerably
suppressed on the proton and should mainly occur on the neutron~\cite{max}.

Several attempts to explain the observed bump have been recently done.
The authors of \cite{tia,tia1,kim} suggested that a narrow $P_{11}(1670)$
resonance can produce the observed structure in the neutron cross section. The inclusion of such resonance
into the modified version of an isobar model
for $\eta$ photoproduction $\eta$-MAID\cite{maid}
generates a narrow peak in the cross section
on the free neutron.  The peak is transformed into
a wider bump similar to experimental observation if Fermi motion
is taken into account~\cite{tia,tia1}. Alternatively, the authors of
\cite{skl} have demonstrated that
the bump in the $\gamma n \to \eta n$ cross section could be explained
in terms of photoexcitation of the $S_{11}(1650)$ and $P_{11}(1710)$ resonances.
In both \cite{tia} and \cite{skl} the angular dependence has been
suggested as a benchmark signature.

Any decisive conclusion about the nature of
the anomalous behavior of the neutron cross section
requires the comparison of theoretical models
and experimental data. This procedure is ambiguous:
model calculations are usually performed for the free neutron
while measured quasi-free observables are smeared by
Fermi motion. The significant influence of Fermi motion on
differential cross sections is shown in \cite{tia}.
Furthermore, quasi-free cross sections are distorted due
to re-scattering and final-state interaction (FSI).
Those events which originate from re-scattering and FSI are in part
eliminated in data analysis. Accordingly the measured cross section
might be smaller than the cross section on the free neutron smeared by Fermi motion.

Free-proton data would provide more clear information.
If photoexcitation of any resonance occur on the neutron
it can occurs also on the proton even being suppressed
by any reasons. For example, the exact $SU(3)$ would forbid
the photoexcitation on the non-strange pentaquark from the anti-decuplet on the proton.
However accounting for the $SU(3)$
violation leads to the cross section of its photoproduction on the
proton 10-30 times smaller (but not 0) than that on the neutron~\cite{max,yang}.

The $\eta$ photoproduction on the proton below $W\sim 1.7$ GeV
is dominated by photoexcitation of the $S_{11}(1535)$ resonance.
This resonance contributes to only the $E_{0}^{+}$ multipole.
$|E_{0}^{+}|^2$ is the major component of the cross section
\vspace*{-0.2cm}
\begin{equation}
\sigma \sim |E_{0}^{+}|^2+\text{interference terms}
\end{equation}
while other multipoles contribute through the interference
with $E_{0}^+$ or between themselves.
A narrow weakly-photoexcited state with the mass below $1.7$ GeV would
appear in the cross section as a small peak/dip structure
on the slope of the dominating $S_{11}(1535)$ resonance.
In experiment this structure would be in addition smeared by
the resolution of a tagging system (for example, the resolution of
the tagging system at GRAAL is 16 MeV FWHM), and might be hidden
due to inappropriate binning.

Polarization observable - the polarized photon beam asymmetry $\Sigma$
is much less affected by the $S_{11}(1535)$ resonance. This observable
is the measure of azimuthal anisotropy of a reaction yield relatively
the linear polarization of the incoming photon.
In terms of $L\leq 1$ multipoles the expression for the beam asymmetry
does not include the multipole
$E_{0}^{+}$\cite{tab}
\begin{eqnarray}
\Sigma(\theta)\sim \frac{3\sin^2\theta}{2}Re(-3|E_1^+|^2+|M_1^+|^2-\\
\nonumber{-2M_1^{-*}(E_1^+-M_1^+)+2E_1^{+*}M_1^+).}
\end{eqnarray}
This observable is mostly governed by the multipoles others
than $E_0^+$ and therefore is much more sensitive to signals
of non-dominant resonances than the cross section.

\begin{figure}
\vspace*{-0.3cm}
\centerline{\epsfverbosetrue\epsfxsize=8.8cm\epsfysize=6.8cm\epsfbox{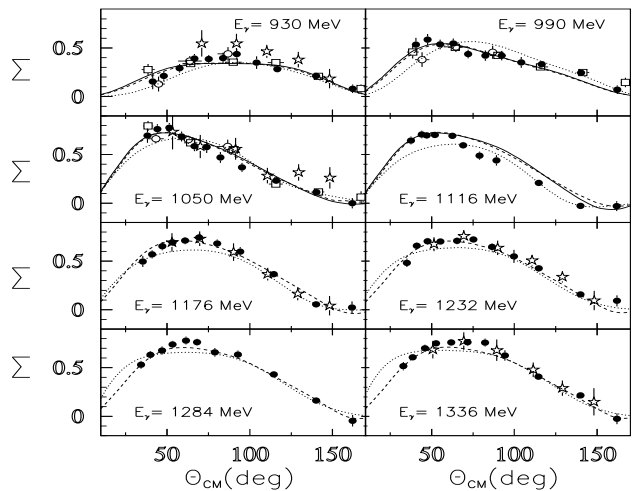}}
\caption{ Published data (in part) for $\Sigma$ beam asymmetry
for $\eta$ photoproduction on the free proton. Open points are from \protect\cite{gra2}:
squares correspond to the detection of two photons from $\eta \to 2\gamma$ decays
in the BGO ball; circles indicate the data obtained detecting one photon in
the forward shower wall and the second in the BGO ball. Black circles are
data from \protect\cite{gra3}. Stars are the results from \protect\cite{bon1}.
Solid lines are  our calculations based on the SAID multipoles only.
Dashed lines are the original E429 solution of the SAID partial wave analysis.
Dotted lines are $\eta$-MAID predictions.}
\vspace*{-0.3cm}
\label{fig:as1}
\vspace*{-0.3cm}
\end{figure}

For $\eta$ photoproduction on the proton the beam asymmetry $\Sigma$
has been measured twice at GRAAL~\cite{pi0}. First results\cite{gra2} covered
the energy range from threshold to $1.05$~GeV.
Two statistically-independent but consistent sets of data points
have been  reported (Fig.~\ref{fig:as1}).
These data sets have been produced using
two different samples of events:\\
i) Events in which two photons from $\eta \to 2\gamma$ decays
were detected in the BGO Ball\cite{bgo};\\
ii) Events in which one of the photons emitted at the angles
$\theta_{lab}\leq 25^{\circ}$ was detected in the forward
shower wall\cite{rw}, and the other in the BGO ball.\\
The second type of events was found to be particularly efficient
at forward angles and energies above $0.9$~GeV.
The results have shown a marked peaking at forward angles and
$E_{\gamma} \sim 1.05$~GeV.

An extension to higher energies up to $1.5$ GeV has been reported in \cite{gra3}.
Two samples of events have been merged and analyzed together.
This made it possible to  reduce significantly error bars
at forward angles and to retrieve a maximum in the angular
dependence at $50^{\circ}$ and $E_{\gamma}\sim 1.05$ GeV (Fig.~\ref{fig:as1}).

Very recently, a new measurement has been done at CB/TAPS@ELSA
using the different technique of the photon-beam polarization,
the coherent bremsstrahlung from a diamond radiator\cite{bon1}.
The results are in a good agreement with \cite{gra2, gra3} but exhibit
slightly larger error bars (Fig.~\ref{fig:as1}).

In all \cite{gra2,gra3,bon1} the main focus was done on the
angular dependencies of the beam asymmetry. Data points have been
produced using relatively narrow angular bins but nearly
60~MeV wide energy bins. So wide energy bins did not allow
to reveal any narrow peculiarities in the energy dependence of
the beam asymmetry.
An ultimate goal of this work was to produce beam asymmetry data using narrow
bins in energy, in order to retrieve in details the dependence of the beam asymmetry
on the photon energy in the region of  $E_{\gamma}=0.85 - 1.15$~GeV
(or $W=1.55 - 1.75$~GeV) and to search for a signal of a narrow resonance.

In this Letter we present the revised analysis of data collected
at the GRAAL facility~\cite{pi0} in 1998 - 1999. Only two
experimental runs are used in the analysis, in order to avoid
additional (up to $\pm 8$ MeV)  uncertainties in the determination of
the photon energy related to different adjustment and
calibration of the GRAAL tagging system in the different run periods.
The data collection had been carried out
as a sequence of alternate measurements with two orthogonal
linear polarization states of a photon beam produced through the backscattering
of laser light on 6.04~GeV electrons circulating in the storage
ring of European Synchrotron Radiation Facility.
The degree of polarization was dependent on photon energy
and varied from 0.5 to 0.85 in the energy range of
$E_{\gamma}=0.85 - 1.15$~GeV.

The procedure of selection of events is similar to
that used in \cite{gra2, gra3}. Two described above types of events
are considered. At the initial stage the first type of events
$\eta$ meson is identified by means of
the invariant mass of two photons detected in the BGO ball
and its momentum is reconstructed from photon
energies and angles. Then measured parameters of
the recoil proton are compared with ones calculated using
kinematics constrains.
Those events in which one of the photons
is detected in the forward shower wall \cite{rw},
are analyzed in a different way: the initial selection
is done using the missing mass calculated from
the energy of the incoming photon and the measured momentum of the recoil proton.

After that a kinematical fit is applied for both types of events.
The center-of-mass angles of $\eta$ and the $\phi$-angles of the reaction plane
are determined by a $\chi^2$ minimization procedure comparing
the calculated energies and angles in the laboratory system with
the measured ones and their estimated errors.
This procedure provides the most accurate determination the reaction
$\theta$ and $\phi$ angles and allows to reduce the influence of 
the detector granularity.
For the second type of events, it is also allows the determination the energy
of the photon detected in the forward wall (not provided by this detector).
After that the events are selected using kinematics constraints
and the value of $\chi^2$.

\begin{figure}
\vspace*{-0.1cm}
\centerline{\epsfverbosetrue\epsfxsize=8.0cm\epsfysize=6.5cm\epsfbox{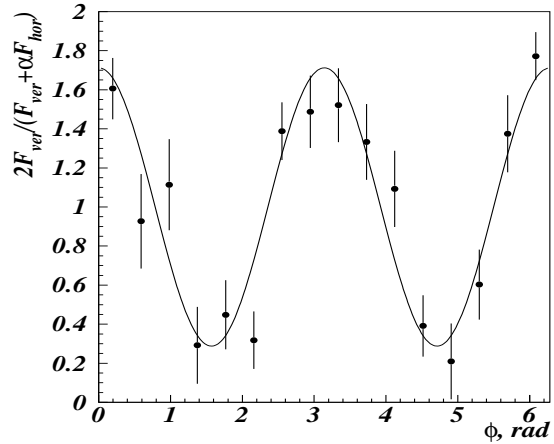}}
\caption{Normalized azimuthal distribution at $E_{\gamma}=1.04~GeV$
 and $\theta_{cm}=43^\circ$. The fitted curve ($1+P\Sigma cos(2\phi)$)
 makes it possible to extract the beam asymmetry $\Sigma$.}
\vspace*{-0.5cm}
\label{fig:as3}
\vspace*{-0.1cm}
\end{figure}

At the final stage both samples of events are merged
and used together to extract beam asymmetries.
For photons linearly polarized in the vertical plane with a
polarization degree $P$, the polarized cross section can be written as
\vspace*{-.3cm}
\begin{equation}
(\frac{d\sigma }{d\Omega })_{pol}(\phi) = (\frac{d\sigma }{d\Omega })_{unpol}(1+P\Sigma cos(2\phi)),
\vspace{-.0cm}
\end{equation}
 \noindent where $\phi$ is the angle between the reaction plane and
the photon polarization and $\Sigma$ is the beam asymmetry. The
cylindrical symmetry of the GRAAL detector provides the azimuthal
distributions of selected events over the entire range 0-360$^\circ$ of $\phi$
angles. The asymmetry is extracted from the distribution
of events for one of the  polarization states, normalized to
the sum  of two azimuthal distributions for both polarization states
which corresponds to an unpolarized beam
\vspace{-.1cm}
\begin{eqnarray}
\frac{\frac{d\sigma }{d\Omega }_{pol}(\phi)}{\frac{d\sigma}{d\Omega}_{unpol}}=
\nonumber{\frac{2F_{ver}(\phi)}{(F_{ver}(\phi) +\alpha F_{hor}(\phi))} = 1+P\Sigma cos(2\phi), }
\vspace{-.0cm}
\end{eqnarray}

\noindent $F_{hor}$ and $F_{ver}$ are the measured azimuthal
distributions of events for each polarization state, $\alpha$ is the
ratio of the beam fluxes corresponding to the vertical and horizontal
polarizations. An example of the extraction of beam asymmetry is
shown in Fig.~\ref{fig:as3}.

\begin{figure}
\vspace*{0.5cm}
\centerline{\epsfverbosetrue\epsfxsize=9.0cm\epsfysize=9.0cm\epsfbox{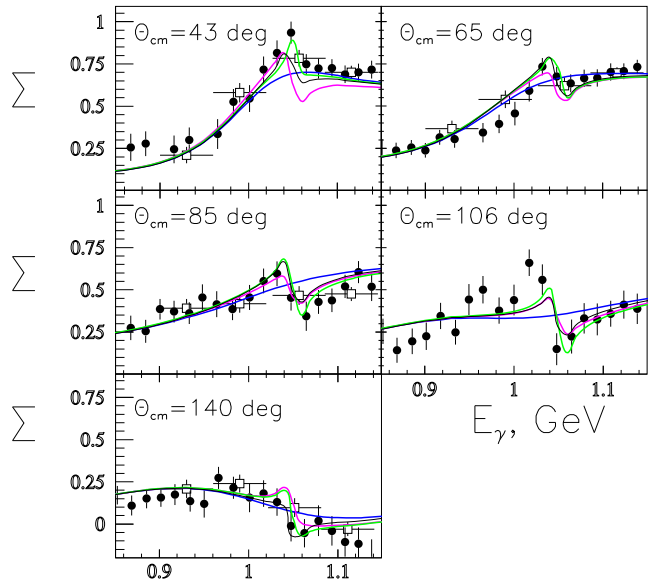}}
\caption{Beam asymmetry $\Sigma$ for $\eta$ photoproduction on the free proton
obtained with narrow energy bins (black circles). Open circles
are data from \protect\cite{gra3}. Blue lines are our calculations
based on the SAID multipoles only, magenta lines are the calculations which
include the $P_{11}(1688)$ resonance with the width $\Gamma=8$ MeV,
green lines are calculations with the $P_{13}(1688)$ ($\Gamma=8$ MeV),
black lines are calcultaions with the $D_{13}(1688)$ ($\Gamma=8$ MeV).
}
\vspace*{-0.5cm}
\label{fig:as2}
\vspace*{-0.3cm}
\end{figure}

The results are shown in Fig.~\ref{fig:as2}.
Data points are obtained
using narrow energy bins $\Delta E_{\gamma} \sim 16$
MeV. Angular bins are chosen to be rather wide, about
$20 - 40^{\circ}$, to gain statistics and hence reduce error bars.

At forward angles $\theta_{cm}=43^{\circ}$ and $E_{\gamma}=1.04$~GeV
data points form a sharp peak with the asymmetry in its maximum reaching
values as large as 0.94. The peak becomes less pronounced but clear at $65^{\circ}$.
It is replaced by an oscillating structure at $85^{\circ}$ and at $105^{\circ}$.
At more backward angles the values of asymmetry above $1.05$~GeV drop down almost
to 0 (Fig.~\ref{fig:as1}) while statistical errors grow up.

Both the peak at forward angles and the oscillating structure at central angles
together exhibit an interference pattern which may be a signal of
a narrow nucleon resonance.
To examine this assumption, we employ the multipoles of the recent E429 solution
of the SAID partial-wave analysis~\cite{str1} for $\eta$ photoproduction
adding to them a narrow resonance in the Breit-Wigner form (see e.g. \cite{maid}).
Our test calculations without an additional resonance well coincide with the original
SAID predictions(Fig.~\ref{fig:as1}), with a minor discrepancy
arising from computational errors.

The narrow $S_{11}$, $P_{11}$, $P_{13}$, and $D_{13}$ resonances are tried one by one.
The contribution of these resonance has been parametrized
by the mass, width, photocouplings (multiplied by square of $\eta N$ branching),
and the phase. These parameters are varied in order to achieve the best
reproduction of experimental data.
The overall $\chi^2$ minimization is avoided.
The narrow-resonance parameters are varied whereas SAID multipoles we keep
fixed. We consider the SAID multipoles  as a good model for the non-resonant
and/or wide resonances contributions.
The difference between calculated and experimental
values of the asymmetry $\Sigma$ after points in the maximum/minimum of
the peak/dip structure and nearby are used as a criterion
for the minimization.

The curves corresponding to the SAID multipoles only
are smooth and do not exhibit any structure (Fig.~\ref{fig:as2}).
The inclusion of either $P_{11}$ or $P_{13}$ or $D_{13}$ resonances
allows to significantly improve the agreement between
data and calculations and to reproduce the peak/dip structure.
The mass of the included resonances is strongly constrained
by experimental data. Its values belong to the range of $M_R=1.686 - 1.690$~GeV.
The best agreement with data have been obtained
with the width of $\Gamma \sim 8$~MeV. However reasonable curves
are obtained in the range of
$\Gamma \leq 15$~MeV.

The $S_{11}$ resonance generates a dip at $43^{\circ}$ in the entire
range of variation of its photocoupling and phase. This indicates that
the observed structures most probably can not be attributed to a narrow
$S_{11}$ resonance.

The curves shown in Fig.~\ref{fig:as2} corresponds to
$\sqrt{Br_{\eta N}} A^p_{1/2}\sim 2\cdot 10^{-3}$~GeV$^{-1/2}$, and
$\sqrt{Br_{\eta N}} A^p_{1/2}\sim -0.3\cdot 10^{-3}$~GeV$^{-1/2}$,
$\sqrt{Br_{\eta N}} A^p_{3/2}\sim 1.7\cdot 10^{-3}$~GeV$^{-1/2}$,
and $\sqrt{Br_{\eta N}} A^p_{1/2}\sim -0.1\cdot 10^{-3}$~GeV$^{-1/2}$,
$\sqrt{Br_{\eta N}} A^p_{3/2}\sim 0.9 \cdot 10^{-3}$~GeV$^{-1/2}$,
for the $P_{11}$, $P_{13}$, and the $D_{13}$ resonances respectively.
The obtained value of $\sqrt{Br_{\eta N}} A^p_{1/2}$ for
the narrow $P_{11}$ resonance is in good agreement with estimates for the non-strange
pentaquark from the antidecuplet performed in Chiral Quark-Soliton
Model~\cite{max,yang}. If to compare the value with the analogous quantity for
the neutron extracted in the phenomenological analysis of the GRAAL and
CB-TAPS data \cite{akps,tia1}, one may deduce
the ratio $A^n_{1/2}/A^p_{1/2}\sim 6-9$. This ratio is close to that expected
for the non-strange pentaquark in the Chiral Quark-Soliton model \cite{max,yang} and
is essentially larger than the one used in \cite{tia1}.
So large ratio of photoproduction amplitudes indicates the strong
suppression of photoexcitation of this resonance on the proton
and explains why it is not (or poorly) seen in the free-proton cross section.

The insertion of a narrow state into fixed multipoles
does not allow to determine the statistical significance of the signal of the
possible resonance and to fix its quantum numbers. For that, one has to vary
not only parameters of the included resonance but also the basic multipoles.
This could be done at such PWA facilities as SAID and MAID.
Furthermore, new experimental data from photon factories equipped with
high-resolution tagging systems are needed for that.

The mass estimate for the underlying resonance
is about 5-10~MeV higher than the value obtained in $\eta$ photoproduction
on the neutron. This could be explained by the mass shift of
a narrow resonances in nuclear medium
\begin{equation}
\Delta M = -\frac{<P_{F}^2>}{2M_{R}} \sim -5 MeV
\end{equation}
where $<P_{F}>$ is the average Fermi momentum in the deuteron,
$M_{R}$ is the mass of the resonance.

In summary we report  the evidence for a narrow resonant structure
in the $\Sigma$ beam asymmetry data for $\eta$ photoproduction
on the free proton. This structure may be a manifestation
of a narrow resonance with the mass $M\sim 1.688\pm 0.002\pm 0.005$~GeV and the width
$\Gamma \leq 15$ MeV. As a candidate we tried a narrow $S_{11}, P_{11}$, $P_{13}$
and $D_{13}$ resonances. Among them, either the $P_{11}$ or
$P_{13}$ or $D_{13}$ resonance improves the description of the data.
The same resonance is observed in $\eta$ photoproduction
on the neutron where its photoexcitation is much stronger.

It is a pleasure to thank the staff of the European Synchrotron Radiation Facility
(Grenoble, France) for stable beam operation during the experimental run.
Special thanks to I. Strakovsky for many discussions, encouragement and help with
SAID data base.
We are thankful to Y.~Azimov, A.~Fix, K.~Goeke,  and L.~Tiator
for many valuable discussions.
P.~Druck is thanked for support in data processing.
The authors appreciate very much voluntary help with the analysis
and continuous interest, and support to this work of
K.~Hildermann, D.~Ivanov, I.~Jermakowa, M.~Oleinik, and N.~Sverdlova.
This work has been supported in part by the Sofja Kowalewskaja Programme
of Alexander von Humboldt Foundation and in part by Korean Research Foundation.

%%%%%%%%%%%%%%%%%%%%%%%%%%%%%%%%%%%%%%%%%%%%%%%%%%%%%%

\end{document}